\documentclass[aps,floats,prl,superscriptaddress,
twocolumn]{revtex4}

\usepackage{amsfonts,amsmath} \usepackage{bm} \usepackage{dcolumn}
\usepackage{epsfig} \usepackage{latexsym}
\usepackage{multirow}

\begin{document}

\title{Hidden thermal structure in Fock space}

\author{Chushun Tian}
\email{ct@itp.ac.cn}
\affiliation{CAS Key Laboratory of Theoretical Physics and Institute
of Theoretical Physics, Chinese Academy of Sciences, Beijing 100190,
China}

\author{Kun Yang}
\email{kunyang@magnet.fsu.edu}
\affiliation{National High Magnetic Field Laboratory and Department of Physics, Florida State University, Tallahassee, FL 32306, USA}

\author{Ping Fang}
\affiliation{Institute for Advanced Study, Tsinghua University, Beijing 100084, China}

\affiliation{CAS Key Laboratory of Theoretical Physics and Institute
of Theoretical Physics, Chinese Academy of Sciences, Beijing 100190,
China}

\author{Hai-Jun Zhou}
\affiliation{CAS Key Laboratory of Theoretical Physics and Institute
of Theoretical Physics, Chinese Academy of Sciences, Beijing 100190,
China}

\author{Jiao Wang}
\affiliation{Department of Physics, Key Laboratory of Low Dimensional Condensed Matter Physics (Department of Education of Fujian Province), and Jiujiang Research Institute, Xiamen University, Xiamen 361005, Fujian, China}

\date{\today}
\begin{abstract}
The emergence of quantum statistical mechanics from individual pure states of closed many-body systems is currently under intensive investigations. While most efforts have been put on the impacts of the {\it direct} interaction (i.e., the usual mutual interaction) between particles, here we study systematically and analytically the impacts of the {\it exchange} interaction, that arises from the particle indistinguishability. We show that this interaction leads an overwhelming number of Fock states to exhibit a structure, that can be resolved only by observables adjusted according to system's dynamical properties and from which thermal distributions emerge. This hidden thermal structure in Fock space is found to be related to the so-called limit shape of random geometric objects in mathematics. The structure enables us to uncover, for both ideal and nonideal Fermi gases, new mechanisms for the emergence of quantum statistical mechanics from individual eigenstates.
\end{abstract}

\maketitle

There have been increasing evidences \cite{Shankar85,Deutsch91,Srednicki94,Zelevinsky95,Tasaki98,Rigol08,Popescu06,Lebowitz06} showing that a closed quantum many-body system can act as its own heat bath, leading to the emergence of equilibrium statistical mechanics from pure states (see Refs.~\cite{Rigol16,Izrailev16,Gogolin16,Deutsch18} for review). Notwithstanding this, the ingredients indispensable for this emergence remain an open problem. It has been shown that the direct interaction between particles, via driving many-body quantum chaos, gives rise to complex structures of eigenstates, from which the Fermi-Dirac (FD) and Bose-Einstein (BE) distributions arise \cite{Srednicki94,Izrailev16,Casati98,Izrailev17}. The need of this interaction conforms to standard statistical mechanics \cite{Landau}. Whereas the studies of entanglement entropy suggest that without the direct interaction, the distribution arises also \cite{Yang15,Dhar16,Sarma16,Magan16}. That thermal distributions exist in such a broad range of extreme conditions motivates exploring universal routes to their emergence from pure states. Furthermore, the exchange interaction -- ``a peculiar mutual effect of particles that are in the same quantum state'' \cite{Landau} -- is a common ingredient of quantum many-body systems. It is a building block of standard ensemble-based quantum statistical mechanics, giving rise to the FD and BE distributions and many intriguing phenomena ranging from the BE condensation to the Haldane-Wu statistics in quantum Hall systems \cite{Haldane91,Wu94}. Thus in-depth investigations of the exchange interaction in the emergence of statistical mechanics from pure states are of both fundamental importance and urgent need.

\begin{figure}
\includegraphics[width=8.7cm] {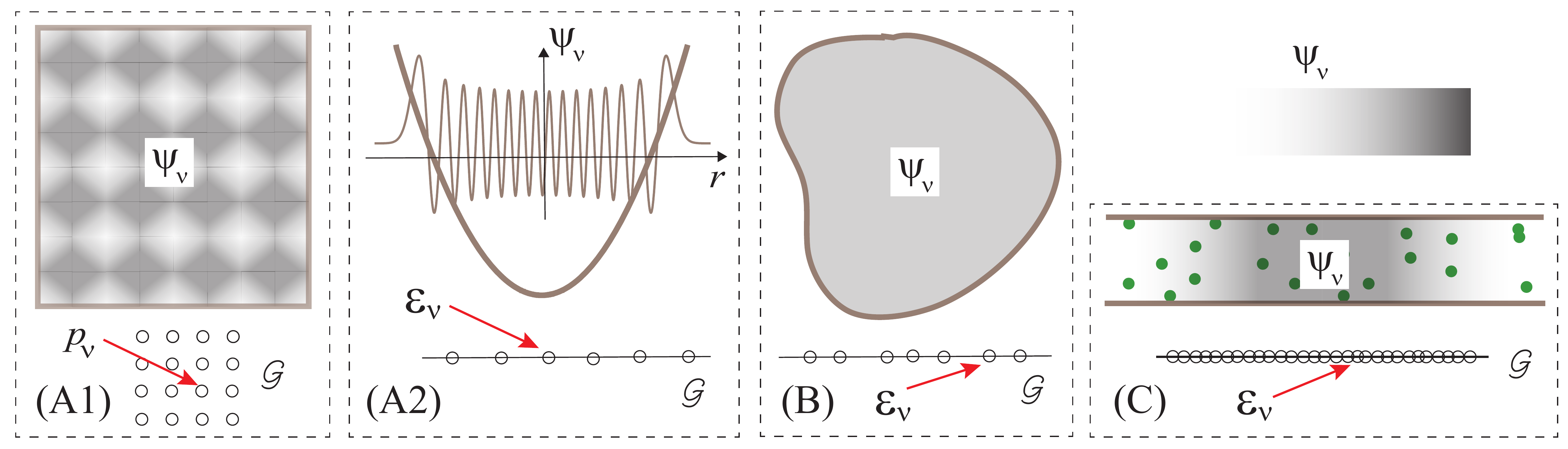}
\caption{Schematic representation of the spatial structures of single-particle eigenstate $\psi_\nu$ and the good quantum number space ${\cal G}$ (empty circles) of Liouville
integrable (A1,2), chaotic (B) and Anderson localized (C) motions in distinct potential constructions. Green dots in (C) are scatterers.}
\label{fig:1}
\end{figure}

In this Letter, for the first time, we study systematically and analytically how the exchange interaction drives thermal equilibrium phenomena at individual pure states. For simplicity we focus on Fermi statistics here. We consider $N$$($$\gg$$1$$)$ indistinguishable fermions confined in a volume \cite{note3} for both situations: with and without the direct interaction. Without the direct interaction an ideal Fermi gas results. In this case the exchange interaction endows the gas with the many-body nature. The eigenstate of this gas is a Fock state $\lambda$, represented by a pattern of the number of particles occupying a single-particle eigenstate. Three classes of representative single-particle eigenstates will be considered, corresponding to distinct single-particle quantum motions (Fig.~\ref{fig:1}): Liouville integrable, chaotic and Anderson localized. To realize the first we put the particle either on a torus (A1) or in a one-dimensional harmonic potential (A2), the second in a chaotic cavity (B), and the third in a quasi one-dimensional cavity which has an infinite extension in one direction and includes random scatterers (C). When the direct interaction is switched on, a nonideal Fermi gas results, whose eigenstate $\Phi$ is a superposition of Fock states. In this work we first uncover a thermal structure hidden in the Fock space, and then study its consequences on both ideal and nonideal Fermi gases.

The main results are summarized in words as follows:
\begin{itemize}
  \item We find that, irrespective of dynamical properties (Liouville integrable, chaotic or Anderson localized) of single-particle quantum motion, for an overwhelming number of Fock states the FD distribution emerges from individual occupation number pattern (cf.~Table~\ref{tab:1}), but can be probed only by appropriate observables. This thermal structure, hidden in individual Fock states, has nothing to do with many-body quantum chaos, but is related to the limit shape of random geometric objects \cite{Okounkov16,Vershik04,Vershik95,Vershik96,Su16}, a subject well explored by mathematicians.
  \item We find that the influence of dynamical properties is to determine whether an observable can resolve the hidden thermal structure. Table~\ref{tab:1} gives the results for the one-particle correlation function $M_{{r}{r}'}$ between two spatial points ${r}$, ${r}'$. It shows that the short-ranged (i.e., small $|$$r$$-$$r'$$|$) correlation is always thermal, implying that if a subsystem is small, an individual $\lambda$, namely, a many-body eigenstate of ideal Fermi gas, acts as the heat bath of the subsystem, irrespective of dynamical properties. This is in spirit consistent with the results for the reduced density matrix based on the {\it canonical typicality} \cite{Tasaki98,Lebowitz06,Popescu06}, which makes no reference to system's constructions. Whereas the long-ranged (i.e., large $|$$r$$-$$r'$$|$) correlation is thermal only if the single-particle quantum motion is chaotic. In this case $\lambda$ acts as the heat bath of the entire system.
  \item We find that, without Berry's conjecture \cite{Srednicki94,Berry77}, the eigenstate $\Phi$ of nonideal Fermi gases on a torus exhibits {\it eigenstate thermalization} \cite{Srednicki94}. Specifically, we show [Eq.~(\ref{eq:19})] that the expectation value of short-ranged spatial correlation at $\Phi$ is thermal: it is governed by the FD distribution, but not the details of the constructions of $\Phi$.
\end{itemize}
\noindent Our findings suggest that the thermal structure hidden in the Fock space, arising from the exchange interaction namely the particle indistinguishability, is a basis of the emergence of various thermal equilibrium phenomena from pure states. In particular, they indicate new mechanisms for the eigenstate thermalization.

{\it Observable-resolved structure $\Lambda$$($$\lambda$$)$ of individual $\lambda$} -- As
mentioned before an individual Fock state $\lambda$ is a
pattern $\{$$n_\nu$$\}$, where $n_\nu$$($$=$$0$,$1$$)$ is the occupation
number at single-particle eigenstate $\psi_\nu$. Here $\nu$ denotes the complete
set of good quantum numbers associated with the single-particle motion, which refers to the eigenmomentum ${p}_\nu$ for integrable motion (A1) and to the eigenenergy $\varepsilon_\nu$ for integrable (A2), chaotic (B) and Anderson localized (C) motions. Given a system all $\nu$ constitute a space, denoted as ${\cal G}$ (Fig.~\ref{fig:1}).
We will show below (in several steps) that the two issues, namely, the structures of $\{n_\nu\}$ resolved by observables and the emergence of FD distribution from individual eigenstates of (non)ideal Fermi gases, are related closely. In this part we first illustrate how distinct observables resolve the structures of individual $\lambda$ at different scales of ${\cal G}$ (cf.~Table \ref{tab:1}).

\begin{table}
\centering
\caption{Structures of individual Fock state $\lambda$ resolved by short- and long-ranged spatial correlations $M_{rr'}$.}
\begin{tabular}{c|c|c|c|c}
\hline\hline
single-particle& \multicolumn{2}{c|}{short-ranged \cite{note4}}& \multicolumn{2}{c}{long-ranged}\\
\cline{2-5}
eigenstate& structure& expectation& structure& expectation\\
$\psi_\nu$&resolved&value&resolved&value\\
\hline
integrable &thermal& thermal &fine&athermal\\
(A1,2)&structure&[Eq.~(\ref{eq:17})]&structure&[Eq.~(\ref{eq:6})]\\
\hline
chaotic &thermal& thermal &thermal&thermal\\
(B)&structure&[Eq.~(\ref{eq:17})]&structure&[Eq.~(\ref{eq:17})]\\
\hline
localized &thermal& thermal &fine&athermal\\
(C)&structure&[Eq.~(\ref{eq:15})]&structure&[Eq.~(\ref{eq:6})]\\
\hline\hline
\end{tabular}
\label{tab:1}
\end{table}

For simplicity we focus on a family of basic observables, namely, the one-particle correlation function $M_{rr'}$ at different ranges of $|$$r$$-$$r'$$|$. This correlation function is the expectation value of $c^\dagger_{{r}'}$$c_{{r}}$, where $c_{{r}}$ ($c^\dagger_{{r}}$) is the annihilation (creation)
operator at ${r}$, at given $\lambda$:
\begin{equation}\label{eq:6}
  M_{{r}{r}'}\equiv\langle \lambda| c^\dagger_{{r}'}c_{{r}}|\lambda\rangle=\sum_{\nu}n_{\nu} C_{\nu}({r},{r}').
\end{equation}
Here $C_\nu$$($${r}$$,$${r}'$$)$$\equiv$$\psi_{\nu}$$($${r}$$)$$\psi_\nu^*$$($${r}'$$)$ is the autocorrelation function of the single-particle eigenstate $\psi_\nu({r})$.

\begin{figure}
\includegraphics[width=8.7cm] {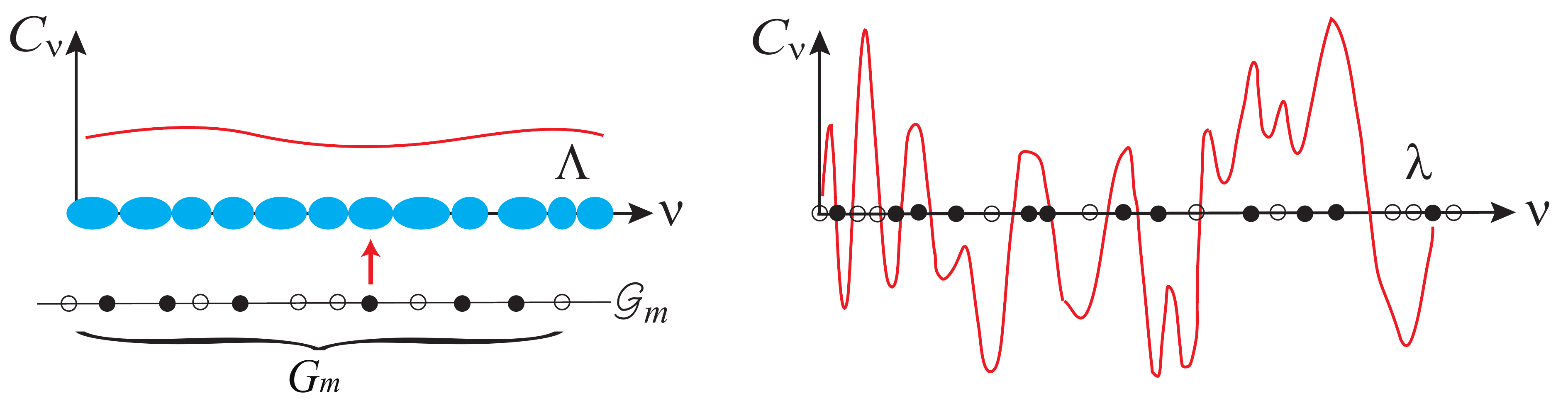}
\caption{Left: When $C_\nu$ varies slowly with $\nu$, a natural decomposition of the space ${\cal G}$ into subspaces ${\cal G}_{m}$ (blue cells) results. Consequently, $M_{rr'}$ can resolve only the structure $\Lambda$ less fine than $\lambda$. Right: When $C_\nu$ varies rapidly with $\nu$, the decomposition does not follow and $M_{rr'}$ can resolve the fine structure of $\lambda$. Solid (empty) circles denote (un)occupied eigenstates $\nu$.}
\label{fig:3}
\end{figure}

(i) If $C_{\nu}$$($${r}$$,$${r}'$$)$ varies slowly with $\nu$ (Fig.~\ref{fig:3}, left), i.e.,
\begin{equation}\label{eq:10}
    C_{\nu}({r},{r}')\approx C_{\nu'}({r},{r}'),\quad {\rm for\, nearest}\, \nu,\nu',
\end{equation}
(we will justify this equation and derive the corresponding conditions for distinct dynamical systems later.) then ${\cal G}$ has a ``natural'' decomposition into many subspaces ${\cal G}_{m}$. In each ${\cal G}_{m}$, $C_\nu$ ($r$,$r'$ fixed) and $\varepsilon_\nu$ are approximately a constant, denoted as $C_{m}$$($$r$$,$$r'$$)$ and $\varepsilon_m$, respectively, i.e.,
\begin{eqnarray}\label{eq:16}
    \begin{array}{c}
       {\cal G}=\oplus_m {\cal G}_{m},\\
\forall \nu\in {\cal G}_{m}: C_\nu(r,r')\approx C_{m}(r,r'),\,\varepsilon_\nu\approx \varepsilon_m.
    \end{array}
\end{eqnarray}
By Eq.~(\ref{eq:10}) the number of elements of ${\cal G}_{m}$, denoted as $G_m$, is $\gg$$1$ \cite{note1}. Using the decomposition (\ref{eq:16}) we obtain:
\begin{equation}\label{eq:14}
      \sum_\nu n_{\nu} C_{\nu}(r,r')=\sum_m C_{m}(r,r')\sum_{\nu\in {\cal G}_{m}}n_{\nu}.
\end{equation}
With the help of this result we reduce Eq.~(\ref{eq:6}) to
\begin{eqnarray}
\label{eq:5}
  M_{{r}{r}'}= {\sum_{m}} N_m C_{m}({r},{r}'),\, N_m=\sum_{\nu\in {\cal G}_{m}}n_{\nu}.
\end{eqnarray}
Therefore, provided that Eq.~(\ref{eq:10}) holds, $M_{{r}{r}'}$ cannot resolve $n_\nu$ at a specific $\nu$; rather, it resolves a less fine structure $\{$$N_m$$\}$$\equiv$$\Lambda$$($$\lambda$$)$, which is constrained by \cite{note2}:
\begin{eqnarray}
\label{eq:21}
\begin{array}{c}
  \sum_mN_m=N\,(=\sum_\nu n_\nu),\\
  \sum_mN_m\varepsilon_{m}\approx E\,(=\sum_\nu n_\nu\varepsilon_\nu).
\end{array}
\end{eqnarray}

(ii) If $C_{\nu}$$($${r}$$,$${r}'$$)$ varies rapidly with $\nu$ (Fig.~\ref{fig:3}, right), then neither the decomposition (\ref{eq:16}) nor the reduction (\ref{eq:5}) follows. As $M_{rr'}$ is given by Eq.~(\ref{eq:6}), a fine tuning in the pattern $\{$$n_\nu$$\}$ can lead to a significant change in $M_{rr'}$. So $M_{rr'}$ can probe the fine structure of $\{$$n_\nu$$\}$.

Here we make two remarks. First, the decomposition (\ref{eq:16}) resembles some ideas of von Neumann \cite{von Neumann29} regarding the fundamentals of statistical mechanics of closed quantum systems. To be specific, that observables can induce the decomposition of the space of quantum states is a basis of von Neumann's analysis \cite{von Neumann29}. However, his decomposition refers to the Hilbert space spanned by the eigenstates of the entire system, which are $\lambda$ ($\Phi$) for (non)ideal Fermi gas. Whereas the decomposition (\ref{eq:16}) refers to ${\cal G}$. Secondly, although $\Lambda$ looks similar to the ``macroscopic state'' of Landau \cite{note10}, there are conceptual differences. Notably, as discussed above $\Lambda$ is resolved only by proper observables, whereas the macroscopic state is independent of observables.

{\it Emergence of thermal structures from $\Lambda(\lambda)$} -- A question naturally is: What does the structure $\Lambda(\lambda)$ look like? To study this problem we note that by definition of $\Lambda$, distinct
$\lambda$ [constrained by Eq.~(\ref{eq:21})] can correspond to the same structure $\Lambda$. The number of $\lambda$ corresponding to $\Lambda$ is given by $\prod_m$$\frac{G_m!}{N_m!(G_m-N_m)!}$$\equiv$$W$$[$$\Lambda$$]$. From this expression we see that $W$ has a sharp peak at some $\Lambda^*$$\equiv$$\{$$N_m^*$$\}$. Physically, this means that an overwhelming number of $\lambda$ have the same observable-resolved structure $\Lambda^*$.

Now we come to show that the thermodynamic relation emerges from an individual $\lambda$ satisfying $\Lambda$$[$$\lambda$$]$$=$$\Lambda^*$: this is in contrast to standard statistical mechanics where thermodynamics is built upon an ensemble. By definition,
\begin{eqnarray}\label{eq:2}
  \frac{\partial S}{\partial N_m}\big|_{\Lambda=\Lambda^*}=\alpha +\beta \varepsilon_{m},\quad S\equiv\ln W[\Lambda].
\end{eqnarray}
Here $\alpha$$,$$\beta$ are the Lagrange multipliers. They depend on $N$$,$$E$, and so do $N_m^*$ and $W$$[$$\Lambda^*$$]$. Taking this and Eqs.~(\ref{eq:21}) and (\ref{eq:2}) into account, we find that
\begin{eqnarray}
  \frac{\partial S}{\partial E}\big|_{\Lambda=\Lambda^*}
\!=\!\frac{\partial}{\partial E}\sum_m N_m(\alpha\!+\!\beta\varepsilon_{m})\big|_{\Lambda=\Lambda^*}\!=\!\beta,
\label{eq:8}
\end{eqnarray}
where in deriving the last equality we have used the fact that $N,E$ are independent variables. Similarly, we have
\begin{eqnarray}
  \frac{\partial S}{\partial N}\big|_{\Lambda=\Lambda^*}
\!=\!\frac{\partial}{\partial N}\sum_m N_m(\alpha\!+\!\beta\varepsilon_{m})\big|_{\Lambda=\Lambda^*}\!=\!\alpha.\,
\label{eq:11}
\end{eqnarray}
Thus $S$$=$$\ln$$W$$[$$\Lambda^*$$]$ gives the thermodynamic entropy \cite{note3}, $\beta$ the inverse thermodynamic temperature $\frac{1}{T}$, and $-$$\frac{\alpha}{\beta}$ the chemical potential $\mu$. So Eqs.~(\ref{eq:8}) and (\ref{eq:11}) reduce to $\frac{\partial S}{\partial E}$$|_{\Lambda=\Lambda^*}$$=$$\frac{1}{T}$
and $\frac{\partial S}{\partial N}$$|_{\Lambda=\Lambda^*}$$=$$-$$\frac{\mu}{T}$.
Note that these relations are independent of the explicit form of $\Lambda^*$.

Furthermore, by substituting the explicit form of $W$$[$$\Lambda$$]$ into Eq.~(\ref{eq:2}) we obtain
\begin{eqnarray}\label{eq:3}
    N_m^*/G_m=(e^{\frac{\varepsilon_m-\mu}{T}}+1)^{-1}\equiv f_{{\rm FD}}(\varepsilon_m).
\end{eqnarray}
So $\Lambda^*$ is determined by the FD distribution $f_{\rm FD}$. Namely, it is a thermal structure. Unlike in standard textbooks \cite{Landau}, here $f_{{\rm FD}}$ refers to individual $\lambda$, instead of an ensemble. In fact, the emergence of FD (BE) distribution from pure states has recently appeared as a new fundamental aspect of statistical mechanics \cite{Dhar16,Yang15,Izrailev16,Izrailev17}. Most importantly, here the thermal structure $\Lambda^*$ namely $f_{\rm FD}$ can be resolved only if Eq.~(\ref{eq:10}) holds, whereas in textbooks of statistical mechanics thermal distributions have nothing to do with observables.

{\it Probing hidden thermal structure $\Lambda^*$} -- Let the Fock space satisfying Eq.~(\ref{eq:21}) be equipped with a uniform probability measure, and $\lambda$ be drawn randomly from this measure. Equation (\ref{eq:5}) and the above analysis then suggest that $M_{{r}{r}'}$ has a {\it typical} value with respect to this measure, because an overwhelming number of $\lambda$ satisfy $\Lambda$$($$\lambda$$)$$=$$\Lambda^*$. Combining Eqs.~(\ref{eq:5}) and (\ref{eq:3}) we find that, provided that Eq.~(\ref{eq:10}) holds, this typical value is
\begin{equation}\label{eq:17}
  M_{{r}{r}'}=\int d\mu(\nu)(e^{\frac{\varepsilon_{\nu}-\mu}{T}}+ 1)^{-1} C_{\nu}({r},{r}').
\end{equation}
Here $d$$\mu$$($$\nu$$)$ is the number of single-particle eigenstates in the good quantum number interval $d\nu$. In Eq.~(\ref{eq:17}), the left-hand side is the expectation value of $c^\dagger_{{r}'}$$c_{{r}}$ at $\lambda$, while the right-hand side is the thermal average of $C_\nu$. Note that the latter is determined by $T$$,$$\mu$ and thus fine tunings of $\{$$n_\nu$$\}$ do not change the value of $M_{{r}{r}'}$. This implies that, for an overwhelming number of (but not all) $\lambda$, $M_{{r}{r}'}$ takes the same value, which is the onset of the eigenstate thermalization \cite{Rigol08,Srednicki94,Deutsch91} of ideal Fermi gases.

The remainder is to find the conditions for Eq.~(\ref{eq:10}) to hold. Below we consider the four single-particle quantum motions in Fig.~\ref{fig:1}, and show that {\it precisely at this point, dynamical properties make significant differences}. In essence, distinct motions results in distinct spatial structures of $\psi_\nu$ (Fig. 1) and accordingly the autocorrelation $C_\nu(r,r')$ displays distinct dependences on $\nu$.

(A1) The eigenstate $\psi_{\nu}$$($${r}$$)$$=$$\frac{e^{i{p}_\nu\cdot {r}}}{L}$, where $L$ is the torus size. With its substitution into the autocorrelation we obtain $C_{\nu}$$\sim$$ e^{i{p}_\nu\cdot ({r}-{r}')}$. Now we consider the behaviors of this $C_{\nu}$ in the following two cases. (i) $|$${r}$$-$${r}'$$|$$\ll$$L$. In this case since the difference between two nearest neighbors: ${p}_\nu$$,$${p}_{\nu'}$ is ${\cal O}$$($$L^{-1}$$)$, we have $|$$($${p}_\nu$$-$${p}_{\nu'}$$)$$\cdot$$($${r}$$-$${r}'$$)$$|$$\ll$$1$. From this we find that the phase: ${p}_\nu$$\cdot$$($${r}$$-$${r}'$$)$ varies slowly with $\nu$, and justify Eq.~(\ref{eq:10}). Thus we have Eq.~(\ref{eq:17}): the short-ranged $M_{rr'}$ is thermal. (ii) $|$${r}$$-$${r}'$$|$$=$${\cal O}$$($$L$$)$. In this case we have $|$$($${p}_\nu$$-$${p}_{\nu'}$$)$$\cdot$$($${r}$$-$${r}'$$)$$|$$=$${\cal O}$$($$1$$)$ instead. Thus ${p}_\nu$$\cdot$$($${r}$$-$${r}'$$)$ varies rapidly with $\nu$ and Eq.~(\ref{eq:10}) breaks down. So the long-ranged $M_{rr'}$ is given by Eq.~(\ref{eq:6}): it is athermal and cannot be used to probe $\Lambda^*$.

(A2) The eigenvalue $\varepsilon_\nu$$=$$\nu$$+$$\frac{1}{2}$ and corresponding eigenstate $\psi_{\nu}$$($$r$$)$$=$$\frac{\pi^{-1/4}}{\sqrt{2^\nu \nu!}}$$e^{-\frac{r^2}{2}}$$H_\nu$$($$r$$)$,
where $H_\nu$ is the Hermite polynomial. For $N$$\gg$$1$ most fermions occupy highly excited single-particle eigenstates. Thus the sum in Eq.~(\ref{eq:6}) is dominated by large $\nu$, for which $\psi_\nu$$($$r$$)$$\sim$$\cos$$($$\sqrt{2\varepsilon_\nu}$$r$$)$.
Substituting this asymptotic expression into $C_\nu$ and repeating the discussions on (A1), we find that $M_{{r}{r}'}$ is thermal for $|$$r$$-$$r'$$|$$\ll$$\sqrt{E/N}$ and athermal otherwise.

(B) To calculate $C_\nu$ we consider (i) large and (ii) small $\varepsilon_\nu$ separately. For (i) we perform the Wigner transformation: $C_{\nu}$$($${r}$$,$${r}'$$)$$\equiv$$\int$$d$${p}$$e^{-
i({r}-{r}')\cdot {p}}$$\Psi_\nu$$($${q}$$,$${p}$$)$ with ${q}$$\equiv$$\frac{1}{2}$$($${r}$$+$${r}'$$)$, and adopt Berry's conjecture for single-particle chaotic motion \cite{Berry77}: $\Psi_{\nu}$$($${q}$$,$${p}$$)$$=$$\frac{\delta(\varepsilon_\nu-H({q},{p}))}{\int\!\!\int d{q}d{p}\delta(\varepsilon_\nu-H({q},{p}))}$, with $H$ being the single-particle Hamiltonian. This conjecture implies that $\psi_{\nu}({r})$ is homogeneous on large scales. Unlike Ref.~\cite{Srednicki94}, here the conjecture is not regarding many-particle motion. Using the conjecture we obtain $C_\nu$$\sim$$f$$($$\frac{|{r}-{r}'|}{\lambda_{\varepsilon_\nu}}$$)$, with $\lambda_{\varepsilon_\nu}$ being the de Broglie wavelength at energy $\varepsilon_\nu$. The function $f(x)$ oscillates in $x$, whose explicit form is unimportant. For nearest $\nu,\nu'$ and for any ${r}$$,$${r}'$, we have
\begin{eqnarray}
\label{eq:12}
|{r}-{r}'|(\lambda_{\varepsilon_{\nu'}}^{-1}-\lambda_{\varepsilon_\nu}^{-1})
\sim(|{r}-{r}'|/L)(\Delta/\varepsilon_\nu)^{1/2}\ll 1,
\end{eqnarray}
with $\Delta$ being the level spacing and $L$ the cavity size. From this we find that $f$$($$\frac{|{r}-{r}'|}{\lambda_{\varepsilon_\nu}}$$)$ is the same for nearest $\nu$$,$$\nu'$. Thus Eq.~(\ref{eq:10}) is justified. For (ii) we do not expect Berry's conjecture to hold, since it is based on the semiclassical approximation. So Eq.~(\ref{eq:10}) breaks down in general. But, the number of particles occupying low-lying single-particle states is $\ll N$. Thus their contributions to the sum in Eq.~(\ref{eq:6}) are negligible, and the breakdown of Eq.~(\ref{eq:10}) has no effects on $M_{{r}{r}'}$. So both short- and long-ranged $M_{rr'}$ are thermal and Eq.~(\ref{eq:17}) follows.

(C) The Anderson localization \cite{Efetov97,Lifshits88,Abrahams10} implies that the eigenvalues $\{$$\varepsilon_\nu$$\}$ are discrete and dense, and $\psi_\nu$ exhibits exponential localization in the extended (longitudinal) direction of cavity (Fig.~\ref{fig:1}). Moreover, the localization center has a singular dependence on $\nu$: as $\nu$ approaches ${\nu'}$ the distance between localization centers of $\psi_\nu$ and $\psi_{\nu'}$ diverges. In addition, the localization length varies with $\nu$. As a result, (i) if $|$${r}$$-$${r}'|$ is sufficiently large, $C_\nu$ varies rapidly with $\nu$. Thus Eq.~(\ref{eq:10}) breaks down and the long-ranged $M_{rr'}$ is athermal. (ii) For ${r}$, ${r}'$ in the same localization volume, the sum in Eq.~(\ref{eq:6}) is dominated by the subset of $\{$$\varepsilon_\nu$$\}$ corresponding to states localized in this volume. Since the localization volume is effectively a chaotic cavity, we can repeat the analysis of (B), and achieve a result similar to Eq.~(\ref{eq:17}), but with $d$$\mu$ replaced by $d$$\mu_{\rm loc}$ giving the number of eigenstates in a localization volume and the interval $d$$\nu$:
\begin{equation}\label{eq:15}
  M_{{r}{r}'}=\int d\mu_{\rm loc}(\nu)(e^{\frac{\varepsilon_{\nu}-\mu}{T}}+ 1)^{-1} C_{\nu}({r},{r}').
\end{equation}
Here $T$$,$$\mu$ are determined by the total number and total energy of particles in this volume.

{\it New mechanism for eigenstate thermalization in nonideal Fermi gas} -- Now we switch on the direct hard-sphere interaction between particles. For simplicity we consider particles on a torus. This system is essentially the same as what was studied in Ref.~\cite{Srednicki94}. An eigenstate $\Phi$ of this system, corresponding to the eigenenergy $E$, is a superposition of $\lambda\in \Omega_{N,E}$, where $\Omega_{N,E}$ is composed of all $\lambda$ satisfying $\sum_\nu$$n_\nu$$=$$N$ and $\sum_\nu$$n_\nu$$\varepsilon_\nu$$=$$E$ ($\varepsilon_\nu$$=$$\frac{p_\nu^2}{2}$):
\begin{equation}\label{eq:13}
  |\Phi\rangle=\sum_{\lambda\in \Omega_{N,E}} C_{\lambda} |\lambda\rangle,\quad \sum_{\lambda\in \Omega_{N,E}} |C_{\lambda}|^2=1.
\end{equation}
Note that for this system the many-body eigenenergy $E$ is exactly the total kinetic energy, and the direct interaction enters only into the coefficients $C_{\lambda}$. By simple algebra we find the one-particle correlation function at $\Phi$:
\begin{eqnarray}
\label{eq:20}
  \langle\Phi|c^\dagger_{{r}'}c_{{r}}|\Phi\rangle = \sum_{\lambda\in \Omega_{N,E}}|C_\lambda|^2 M_{{r}{r}'}\quad\quad\quad\nonumber\\
  + \sum_{\lambda\in \Omega_{N,E}}C_\lambda^*C_{\lambda'}\sum_{\nu\neq\nu'}
\frac{e^{i({p}_\nu\cdot {r}-{p}_{\nu'}\cdot{r}')}}{L^2}\langle\lambda|c_{\nu}^\dagger c_{\nu'}|\lambda\rangle,\quad
\end{eqnarray}
where $M_{{r}{r}'}$$=$$\frac{1}{L^2}$$\sum_\nu$$e^{i{p}_\nu\cdot ({r}-{r}')}$$\langle\lambda|c_{\nu}^\dagger c_\nu|\lambda\rangle$. The left-hand side on Eq.~(\ref{eq:20}) must be translationally invariant, i.e., depend on ${r}$, ${r}'$ via ${r}$$-$${r}'$, for the system has the translation symmetry. But this invariance is violated by the second term on the right-hand side. Thus this term must vanish, giving $\langle$$\Phi$$|$$c^\dagger_{{r}'}$$c_{{r}}$$|$$\Phi$$\rangle$$=$$\sum_{\lambda\in \Omega_{N,E}}$$|$$C_\lambda$$|^2$$M_{{r}{r}'}$. On the other hand, as shown before $W$$[$$\Lambda$$]$ has a sharp peak at $\Lambda^*$. Thus most weights of $|C_{\lambda}|^2$ go to $\lambda$ satisfying $\Lambda$$($$\lambda$$)$$=$$\Lambda^*$ \cite{note7}. For these $\lambda$ and corresponding $M_{{r}{r}'}$ we use the results for (A1) summarized in Table~\ref{tab:1}. In particular, $M_{{r}{r}'}$ takes the typical value (\ref{eq:17}) for $|$$r$$-$$r'$$|$$\ll$$L$, with $T$$,$$\mu$ in Eq.~(\ref{eq:17}) determined by $N$$,$$E$. As $M_{{r}{r}'}$ are insensitive to the fine structure of $\lambda$, it can be further pulled out of the sum: $\sum_{\lambda\in \Omega_{N,E}}$$($...$)$$M_{{r}{r}'}$, giving
\begin{eqnarray}\label{eq:19}
  \langle \Phi| c^\dagger_{{r}'}c_{{r}}|\Phi\rangle&=&M_{{r}{r}'}\sum_{\lambda\in \Omega_{N,E}}|C_\lambda|^2= M_{{r}{r}'}\nonumber\\
  &=&\int\!\!\frac{d{{p}}}{(2\pi)^2}\frac{e^{i{p}\cdot ({r}-{r}')}}{e^{({p}^2/2-\mu)/T}+ 1}.
\end{eqnarray}
Thus the eigenstate thermalization is justified for short-ranged one-particle correlation at both low- and high-lying $\Phi$. Note that unlike Ref.~\cite{Srednicki94}, here we did not use Berry's conjecture made for many-particle chaotic motion, which has not yet been proven.

\begin{figure}
\includegraphics[width=8.7cm] {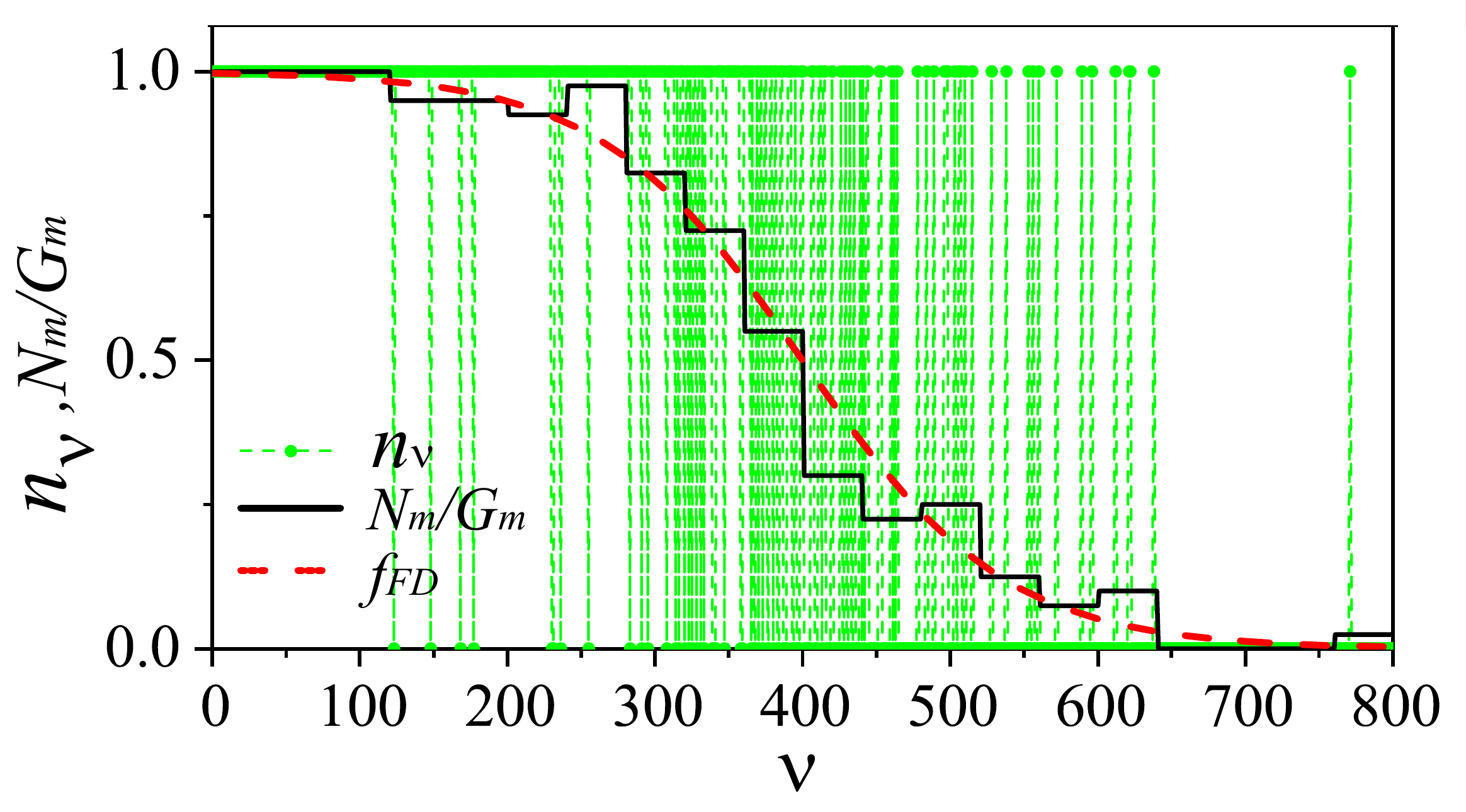}
\caption{A Fock state $\lambda$$=$$\{$$n_\nu$$\}$ of $N$$=$$400$ fermions of total energy $E$$=$$87800$ confined in a harmonic potential, or its number theory equivalent -- a partition of $E$ into $N$ distinct summands, is generated randomly in simulations. For the observable-resolved structure $\Lambda$$=$$\{$$N_m$$\}$ ($G_m$$=$$40$) of a typical $\lambda$, $\{$$N_m$$/$$G_m$$\}$ is fitted well by $f_{\rm FD}$ with $\mu$$=$$400.1$ and $T$$=$$68.84$.}
\label{fig:2}
\end{figure}

{\it Relations between $\Lambda^*$ and the limit shape of random geometric objects} --Finally, we wish to understand the arising of the hidden thermal structure $\Lambda^*$ from more rigorous viewpoints. The point is to view the pattern $\lambda$$=$$\{n_\nu\}$ as a geometric object -- a collection of ``skyscrapers'' located at $\nu$ wherever $n_\nu$$=$$1$ (see, e.g., Fig.~\ref{fig:2}). It turns out that despite the shapes of such objects exhibit randomness, as discovered by mathematicians \cite{Okounkov16,Vershik04,Vershik96,Vershik95,Su16}, when their size is large, upon proper rescaling they can concentrate on a limit shape, which is smooth and nonrandom.

To keep quantitative discussions as simple as possible we consider $N$ indistinguishable free fermions confined in a harmonic potential. The good quantum number space ${\cal G}$ is now the single-particle eigenenergy spectrum, i.e., the set of natural numbers $\mathbb{N}$. (The zero energy is irrelevant for present discussions and thus ignored.) For a Fock state $\lambda$,
$E$$=$$\sum_{\nu=1}^\infty$$\nu$$n_\nu$.
This maps $\lambda$ into a partition of integer $E$ into $N$ distinct summands, a research area for which Euler laid down a foundation \cite{Andrews76}. To be precise, an eigenenergy $\nu$, when its corresponding eigenstate is occupied ($n_\nu$$=$$1$), mimics a summand. In 1941 the field of random integer partitions was opened up \cite{Erdoes41}, and in the past few decades such partitions have been found to bear rich structures \cite{Vershik95,Vershik96,Su16}. (One should not confuse this with the old subject of using standard statistical mechanics to study the number of partitions \cite{Auluck46}.) In particular, the observable
$\varphi_{\lambda}$$($$t$$)$$\equiv$$\int_t^\infty$$\sum_{\nu'=1}^\infty$$n_{\nu'}$$\delta$$($$\nu'$$-$$\nu$$)$$d$$\nu$,
counting at given $\lambda$ the number of summands $\geq$$t$ or, equivalently, the number of particles occupying single-particle eigenstates with eigenenergies $\geq$$t$ \cite{note9}, defines a random stepped curve. As proven rigorously \cite{Vershik96}, this curve has a limit shape. Our results have intimate relations to this.

For illustrations we consider the case where $N$ is not fixed, i.e., $\mu=0$. In this case for sufficiently large $E$, Eq.~(\ref{eq:3}) shows that for an overwhelming number of $\lambda$,
\begin{eqnarray}\label{eq:7}
  \varphi_{\lambda}(t)\stackrel{E\gg 1}{=}\int_t^\infty\frac{d\nu}{e^{\nu/T}+1}=T\ln (1+e^{-t/T}),
\end{eqnarray}
where $T$$=$$\sqrt{12E}$$/$$\pi$ and $E$$=$$\int_0^\infty$$\varphi_\lambda$$($$t$$)$$d$$t$, in agreement with the following theorem \cite{Vershik96}:
\begin{equation*}
  \lim_{E\rightarrow\infty}\mu^E\left\{\lambda: \left|\frac{1}{\sqrt{E}}\varphi_{\lambda}(\sqrt{E}t)+s(t)\right|<\epsilon\right\}=1,\,\forall \epsilon>0.
\end{equation*}
Here $s(t)$ is the Vershik curve: $e^{-\frac{\pi s}{\sqrt{12}}}$$-$$e^{-\frac{\pi t}{\sqrt{12}}}$$=$$1$. Basically, this theorem states that if the set of all partitions $\lambda$ defined above is equipped with a uniform probability measure $\mu^E$, then a typical partition has a limit shape of $-$$s$. Our method, though being less rigorous, has the advantage of being applied to more general conditions (e.g., without the thermodynamic limit) and systems.

Indeed, for generic $N$$,$$E$, for which rigorous results are not available, we have confirmed Eq.~(\ref{eq:3}) numerically. Specifically, we use Monte Carlo method to draw randomly a partition $\lambda$$=$$\{$$n_\nu$$\}$ of $E$ (with $N$ distinct summands) from the uniform probability measure. As shown in Fig.~\ref{fig:2}, a typical $\lambda$ (green pattern), though looks random, has a well-behaved observable-resolved structure $\Lambda$ (stepped curve), in the sense that the pattern $\{$$\frac{N_m}{G_m}$$\}$ concentrates on the smooth distribution $f_{{\rm FD}}$ (red dashed curve) \cite{note1}. ($G_m$ corresponds to the scale over which $C_\nu$ varies and thus to $M_{rr'}$ in specific regimes of $r$$,$$r'$.) Note that here $f_{{\rm FD}}$ is seen to arise not only in an individual $\lambda$, but also for $N$ as small as $400$ (dramatically smaller than $10^{23}$ corresponding to a macroscopic system).

Summarizing, we have shown analytically how the exchange interaction namely the particle indistinguishability gives rise to a hidden thermal structure in the Fock space, and opens up a door to the emergence of thermal equilibrium phenomena from the eigenstates of many-body systems with or without the direct interaction between particles. Furthermore, we have uncovered a new mechanism for the eigenstate thermalization of nonideal Fermi gases on a torus [Fig.~\ref{fig:1}(A1)]. It is natural to generalize this result to nonideal Fermi gases in an Anderson localized cavity [Fig.~\ref{fig:1}(C)]. This issue is currently under investigations. We expect the outcomes to shed new light on the many-body localization \cite{Altshuler06,Mirlin05,Huse14}, especially in view of that the many-body localization is equivalent to the localization in the Fock space. In this work, we focused on the kinematic aspect. The dynamical aspect, especially the interplay between relaxation and the hidden thermal structure, is an important issue in future studies.

\begin{acknowledgments}
We are grateful to G. Casati, S. Fishman, J. C. Garreau, I. Guarneri, D. Huse, H.-H. Lai, A. Polkovnikov and B. Wu for inspiring discussions at various stages of this work. This work is supported by the National Natural Science Foundation of China (Grants No. 11535011, No. 11747601 and No. 11335006) and the National Science Foundation (Grants No. DMR-1644779 and No. DMR-1442366).
\end{acknowledgments}






\begin{thebibliography}{}

\bibitem{Shankar85} R.V. Jensen and R. Shankar, Phys. Rev. Lett. \textbf{54}, 1879 (1985).

\bibitem{Deutsch91} J. M. Deutsch, Phys. Rev. A \textbf{43}, 2046 (1991).

\bibitem{Srednicki94} M. Srednicki, Phys. Rev. E \textbf{50}, 888 (1994).

\bibitem{Zelevinsky95} V. Zelevinsky, Annu. Rev. Nucl. Part. Sci. \textbf{46}, 237 (1996).

\bibitem{Tasaki98} H. Tasaki, Phys. Rev. Lett. \textbf{80}, 1373 (1998).

\bibitem{Popescu06} S. Popescu, A. J. Short, and, A. Winters, Nat. Phys. \textbf{2}, 754 (2006).

\bibitem{Lebowitz06} S. Goldstein, J. L. Lebowitz, R. Tumulka, and N. Zanghi, Phys. Rev. Lett. \textbf{96}, 050403 (2006).

\bibitem{Rigol08} M. Rigol, V. Dunjko, and M. Olshanii, Nature \textbf{452}, 854 (2008).

\bibitem{Rigol16} L. D'Alessio, Y. Kafri, A. Polkovnikov, and M. Rigol, Adv. Phys. \textbf{65}, 239 (2016).

\bibitem{Izrailev16} F. Borgonovi, F. M. Izrailev, L. F. Santos, and V. G. Zelevinsky, Phys. Rep. \textbf{626}, 1 (2016).

\bibitem{Gogolin16} C. Gogolin and J. Eisert, Rep. Prog. Phys. \textbf{79}, 056001 (2016).

\bibitem{Deutsch18} J. M. Deutsch, Rep. Prog. Phys. \textbf{81}, 082001 (2018).

\bibitem{Casati98} F. Borgonovi, I. Guarneri, F. M. Izrailev, and G. Casati, Phys. Lett. A \textbf{247}, 140 (1998).

\bibitem{Izrailev17} F. Borgonovi, F. Mattiotti, and F. M. Izrailev, Phys. Rev. E \textbf{95}, 042135 (2017).

\bibitem{Landau} L. D. Landau and E. M. Lifshitz, {\it Statistical physics}, Part 1, 3rd Ed. (Butterworth-Heinemann, Oxford, UK, 1980).


\bibitem{Yang15} H. H. Lai and K. Yang, Phys. Rev. B \textbf{98}, 081110(R) (2015).

\bibitem{Dhar16} S. Nandy, A. Sen, A. Das, and A. Dhar, Phys. Rev. B \textbf{94}, 245131 (2016).

\bibitem{Sarma16} X. Li, J. Pixley, D.-L. Deng, S. Ganeshan, and S. D. Sarma, Phys. Rev. B \textbf{93}, 184204 (2016).

\bibitem{Magan16} J. M. Mag$\acute{\rm a}$n, Phys. Rev. Lett. \textbf{116}, 030401 (2016).

\bibitem{Haldane91} F. D. M. Haldane, Phys. Rev. Lett. \textbf{66}, 1529 (1991).

\bibitem{Wu94} Y.-S. Wu, Phys. Rev. Lett. \textbf{73}, 922 (1994).

\bibitem{note3} The particle mass, the Boltzmann constant and the Planck constant are all set to unity throughout this work.


\bibitem{Okounkov16} A. Okounkov, Bull. Amer. Math. Soc. \textbf{53}, 187 (2016).

\bibitem{Vershik04} A. M. Vershik, J. Math. Sci. \textbf{119}, 165 (2004).

\bibitem{Vershik95} A. M. Vershik, {\it Proceedings of the International
Congress of Mathematicians, Z{\" u}rich, 1994} (Birkh{\"a}user, Basel, 1995).

\bibitem{Vershik96} A. M. Vershik, Funct. Anal. Appl. \textbf{30}, 90 (1996).

\bibitem{Su16} For a recent review, see, e.g., Z. G. Su, Adv. Math. \textbf{45}, 861 (2016).

\bibitem{note4} For (A2) and (B) it is further required that $r,r'$ are deep in the bulk.

\bibitem{Berry77} M. V. Berry, J. Phys. A: Math. Gen. \textbf{10}, 2083 (1977).


\bibitem{note1} The exact locations of the boundary between two nearest ${\cal G}_m$, i.e., the values of $G_m$, are unimportant. Small changes in these values have no physical consequences. This was confirmed by numerical experiments on random partition of integers. Specifically, we have changed $G_m$ for the partition shown in Fig.~\ref{fig:2} and found that $\{$$N_m$$/$$G_m$$\}$ is fitted well by $f_{{\rm FD}}$, with the parameters $T$ and $\mu$ being the same as those given in the caption of Fig.~\ref{fig:2}.

\bibitem{note2} In fact, for a quantum gas on a $d$-dimensional torus we have $d$ more constraints, imposed by the conservation law of total momentum. However, these additional constraints are not essential for this system. Indeed, if we introduce $d$ Lagrange multipliers: $($$u_1$$,$$u_2$$,$$\cdots$$,$$u_d$$)$$\equiv$${u}$ corresponding to these constraints, the distribution (\ref{eq:3}) is modified to be $($$e^{({p}^2/2-{u}\cdot {p}-\mu)/T}$$+$$1$$)^{-1}$, which is similar to the generalized Gibbs ensemble \cite{Rigol16}. This modified distribution can be rewritten as $($$e^{(({p}-{u})^2/2-\tilde\mu)/T}$$+$$1$$)^{-1}$, where $\tilde \mu$$=$$\mu$$+$$\frac{1}{2}{u}^2$. From this we see that the only physical consequence introduced by the additional constraints is that the torus moves in a velocity ${u}$ (with respect to the laboratory reference frame). However, we can always adjust the reference frame in which the torus is at rest.

\bibitem{von Neumann29} J. von Neumann, Z. Phys. \textbf{57}, 30 (1929).


\bibitem{note10} See $\S 40$ of Ref.~\cite{Landau}: ``Any macroscopic state of an ideal gas may be described as follows. Let us distribute all the quantum states of an individual particle of the gas among groups each containing neighbouring states (which, in particular, have neighbouring energy values), both the number of states in each group and the number of particles in these states being still very large.''

\bibitem{Efetov97} K. B. Efetov, {\it Supersymmetry in disorder and chaos} (Cambridge University, Cambridge, UK, 1997).

\bibitem{Lifshits88} I. M. Lifshits, S. A. Gredeskul, and L. A. Pastur, {\it Introduction to the theory of disordered systems} (Wiley, New York, 1988).

\bibitem{Abrahams10} E. Abrahams, {\it Fifty years of Anderson localization} (World Scientific, Singapore, 2010).

\bibitem{note7} There are rare $\lambda$ which have no thermal structures, i.e., $\Lambda(\lambda)\neq \Lambda^*$. Provided that most weights of $|C_\lambda|^2$ go to these $\lambda$, Eq.~(\ref{eq:19}) ceases to work. But on physical grounds this seems impossible.

\bibitem{Andrews76} G. E. Andrews, {\it The theory of partitions} (Addison-Wesley, London-Amsterdam, 1976).

\bibitem{Erdoes41} P. Erd{\"o}s and J. Lehner,
Duke Math. J. \textbf{8}, 335
(1941).

\bibitem{Auluck46} F. C. Auluck and D. S. Kothari, Math. Proc. Cambridge Phil. Soc. \textbf{42}, 272
(1946).

\bibitem{note9} This observable corresponds to the replacement of $C_\nu$ in Eq.~(\ref{eq:6}) by $\theta(\nu-t)$ [$\theta(x)$ the Heaviside function], which obeys Eq.~(\ref{eq:10}).

\bibitem{Altshuler06} D. Basko, I. Aleiner, and B. L. Altshuler, Ann. Phys. \textbf{321}, 1126 (2006).

\bibitem{Mirlin05} I. V. Gornyi, A. D. Mirlin, and D. G. Polyakov, Phys. Rev. Lett. \textbf{95}, 206603 (2005).

\bibitem{Huse14} R. Nandkishore and D. A. Huse, Ann. Rev. Cond. Mat. Phys. \textbf{6}, 15 (2015).


\end{thebibliography}
\end{document}